\documentclass{arxivIEEEcsmag}

\usepackage[colorlinks,urlcolor=blue,linkcolor=blue,citecolor=blue]{hyperref}

\usepackage[percent]{overpic}

\jvol{XX}
\jnum{XX}
\paper{8}
\jmonth{}
\jname{}
\pubyear{2022}

\makeatletter
\def\ps@plain{}
\makeatother

\begin{document}

\sptitle{}
\editor{}

\title{Overcoming Challenges to Continuous Integration in HPC\vspace{-.5em}}

\author{Todd Gamblin}
\affil{{Lawrence Livermore National Laboratory}}

\author{Daniel S. Katz}
\affil{{NCSA \& CS \& ECE \& iSchool, University of Illinois at Urbana-Champaign}}

\markboth{}{Continuous Integration Challenges for HPC}

\begin{abstract}
 Continuous integration (CI) has become a ubiquitous practice in modern software development, with major code hosting services offering free automation on popular platforms. CI offers major benefits, as it enables detecting bugs in code prior to committing changes.
 While high-performance computing (HPC) research relies heavily on software, HPC machines are not considered ``common'' platforms. This presents several challenges that hinder the adoption of CI in HPC environments, making it difficult to maintain bug-free HPC projects, and resulting in adverse effects on the research community.
In this article, we explore the challenges that impede HPC CI, such as hardware diversity, security, isolation, administrative policies, and non-standard authentication, environments, and job submission mechanisms.
We propose several solutions that could enhance the quality of HPC software and the experience of developers. Implementing these solutions would require significant changes at HPC centers, but if these changes are made, it would ultimately enable faster and better science.
\end{abstract}

\maketitle

\begin{figure*}
  \begin{overpic}[width=\textwidth]{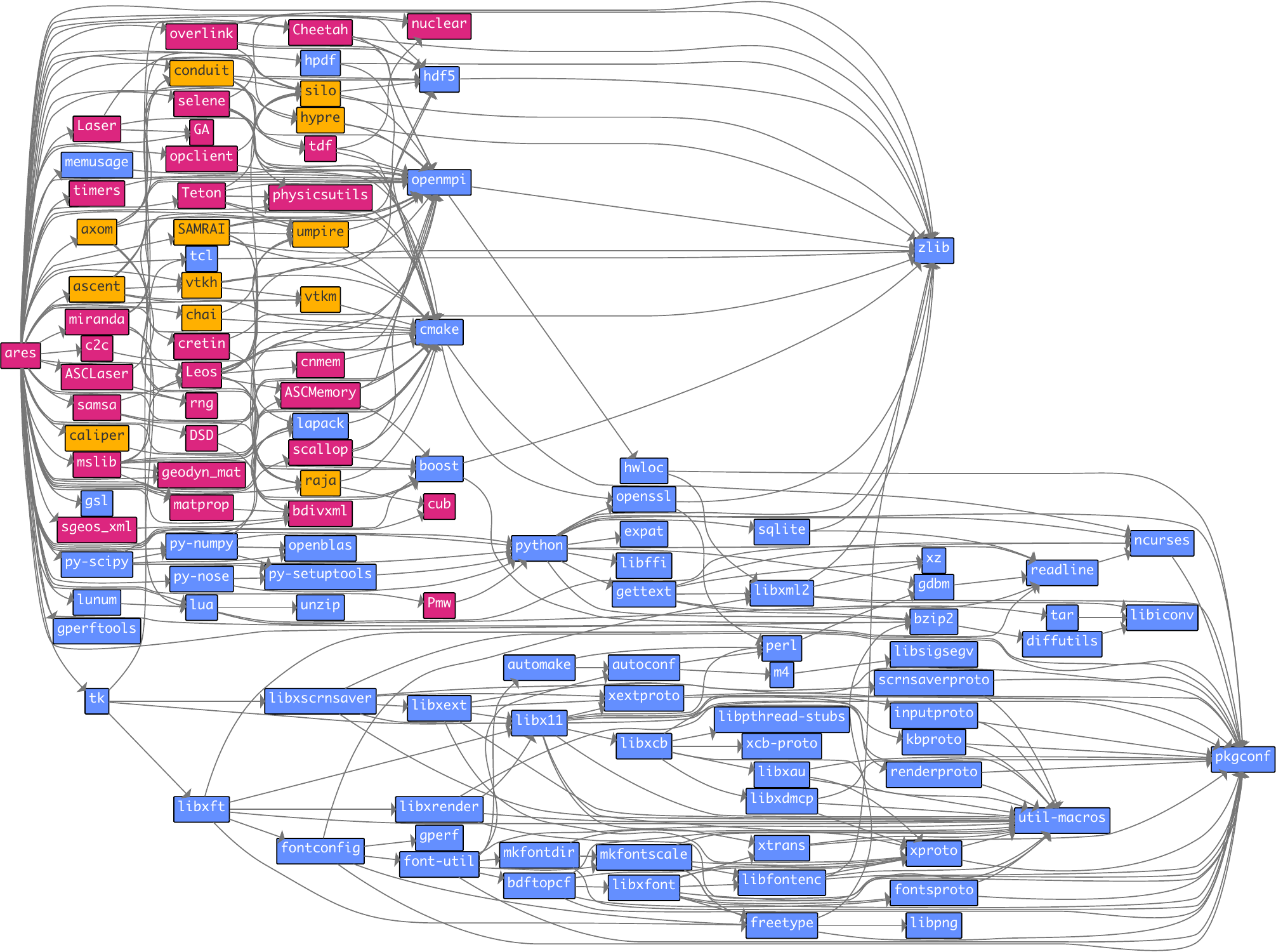}
     \put(75,60){\includegraphics[width=.25\textwidth]{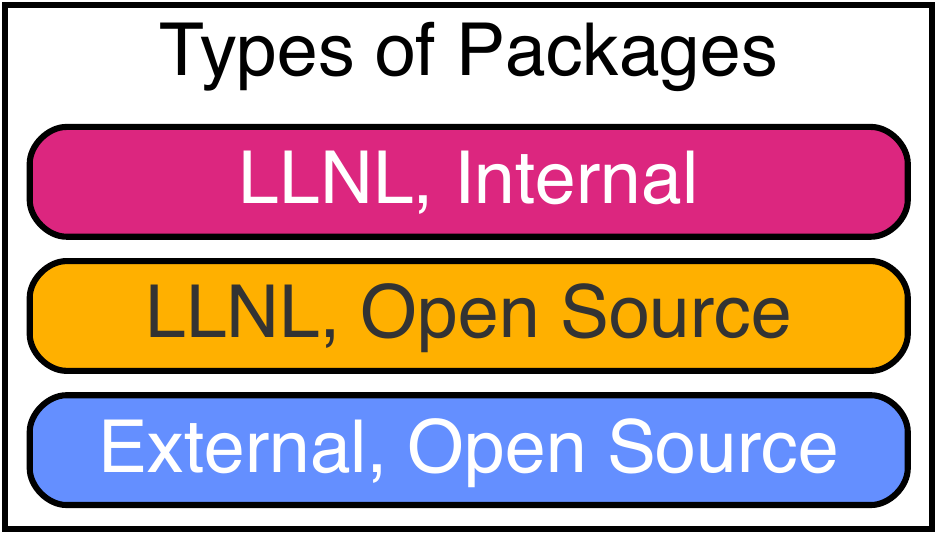}}
  \end{overpic}
  \caption{Dependencies of the ARES multi-physics code:
    31 are internal proprietary packages, 13 are open-source packages developed at LLNL,
    and together these rely on 72 external open-source software packages.
    \label{fig:ares-oss}
  }
\end{figure*}

\chapterinitial{High performance computing} is a key enabler for developing scientific understanding and knowledge.
``High performance'' typically refers to computing that requires large-scale
resources, e.g., those on the Top500 list of the world's fastest machines~\cite{top500}.
HPC sites range from universities with smaller clusters of commodity machines to large,
GPU-accelerated supercomputers at national computing facilities.

HPC systems need application software to be useful. Since around the 1940s, HPC
applications have spanned the computational science domains: simulations and modeling in
climate, physics, chemistry, engineering, etc. More recently, the field grew
to include applications in data analysis and machine learning.
All of these applications rely on other software, from operating systems to libraries
(e.g., for communications and math), and still more software is needed in the development
process: compilation, testing, packaging, and distribution.

Historically, staff at HPC sites developed their own applications, with the vendor of the
HPC system providing the operating system, compilers, and math libraries. Export controls and
other data sensitivity concerns limit access to a large number of HPC applications.
Because of these and more general security concerns, HPC sites only grant access to a set
of known account holders. However, a large fraction of
today's software is developed on social coding platforms like GitHub and GitLab, which
allow a community to perform collaborative planning, development, maintenance, and
testing. These sites not only provide infrastructure for working on code, but offer a large
number of free cloud CPU cycles for continuous integration (CI). Under the CI model,
tests run when developers suggest changes, and the tests ensure that code is
correct {\it before} it is accepted. Developers can thus have high confidence that
the code will work correctly.

While continuous integration is standard practice for developers who can use common
cloud environments, HPC environments introduce challenges to this practice. Technical,
security, and political issues all make it extremely difficult to integrate externally
developed open-source software with internal applications and machines. Even though many
HPC software projects are developed in the open, they must run on closed HPC resources, and it is
increasingly difficult to ensure that the vast majority of modern {\it open-source}
applications will run reliably on HPC systems.

\section{MODERN SOFTWARE IS COMPLEX}

Modern software applications are not monolithic; they integrate packages written by many
authors on different project teams, and they rely heavily on publicly available open-source software. Figure~\ref{fig:ares-oss} shows the software packages used by ARES, a
proprietary multi-physics application used on HPC machines at Lawrence Livermore
National Laboratory (LLNL). These packages include core scientific libraries and
utility libraries for logging, math, I/O, programming models, performance portability,
memory management, and other purposes. A number of build-time
dependencies like compilers and testing frameworks are also included.

Even though 30 core components of Ares are LLNL-proprietary, the other 85 packages are
open source. Of these, 12 are publicly developed by LLNL on GitHub, and the remaining 71
packages are open-source packages developed by others. This situation is not unusual;
{\it most} modern software leverages and depends on open-source components.
Reimplementing all the capabilities provided by the modern open-source-software ecosystem would be impossible or at least impractically expensive for a single organization

Software reuse comes with a cost: integration complexity.
In a project developed by a single team, developers commit code to a common repository, maintaining project consistency. s
In large integrated systems, however, different teams may work on individual components, and developers are responsible for ensuring that all versions of the components are compatible.
Unfortunately, most open-source developers lack access to HPC resources, and even if they have it, many lack the time to manually test their packages in HPC-like environments. HPC developers who leverage open-source software must be prepared to perform extensive porting and integration testing to ensure that the open components work seamlessly on closed systems.

\section{HPC SYSTEMS ARE UNIQUE}

HPC systems are typically designed and built to meet specific local requirements, balancing
expected workload characteristics, hardware options (e.g., number and type of CPUs, GPUs,
and other accelerators; internal networking; storage), packaging, cooling, external networking,
energy usage, cost, etc.
Key components of the local software stack are often bespoke for each system.
For example, proprietary MPI implementations like Cray MPI can only run
on Cray systems. In this case, the Cray MPI license disallows inclusion in
containers or other software distributions that can run in the cloud. The same is true
for math libraries and compilers in the Cray environment. Moreover,
filesystem organization is not standardized.  Paths to tools, libraries, home, and
temporary directories are system-dependent.
Finally, authorization and access may be set by site policies that are often developed locally.

\section{ROADBLOCKS TO CI IN HPC}

Open-source developers are now accustomed to widely available compute cycles for
continuous integration. Major code hosting sites
(GitHub, GitLab, Bitbucket) as well as third-party paid services that integrate with
these sites offer free CI services. Developers can attach workflows to their repositories that run tests
concurrently across Linux, macOS, and Windows, and if they need to test in
custom environments they can bring their {\it own} containerized test environments.
In HPC settings, however, a number of hurdles prevent adoption of automated code building and testing.

\subsection{HPC environments are hard to replicate}

Due to ubiquitous cloud computing, CI is the norm outside of HPC. It has never been
easier to set up automated testing in widely-used software environments.
But, as discussed above, HPC environments are, by definition, special.
For example, it is seldom possible to reliably test {\it optimized} CPU builds in
cloud CI, as the fleet of test systems used by cloud virtual machines (VMs) is often heterogeneous and one
cannot request that a test be run on a specific microarchitecture. So far, there is no (free) cloud-based
CI for GPUs.

Testing scientific workflow systems is even harder. Each workflow system is essentially
a distributed application, and testing a workflow system requires access to the job
submission interface. This access can include authentication and authorization from
remote systems, local environments and configurations, batch scheduler parameters, etc.
Because resource managers are used to run the CI system itself, it is
difficult to vary and test system software and resource managers {\it within} the CI system. In
this case, we need to see how the system is set up in practice, and not completely
isolate from it. We also need interfaces and abstractions that allow us to test
that the system works across different schedulers and configurations.
Without the ability to replicate the software environment of popular HPC systems, it is
very difficult to ensure that open-source software will continue working on them.

\subsection{Security challenges}

HPC machines are large, shared computing systems like clouds, and one obvious way to
replicate the HPC environment in cloud CI would be to offer cycles on a local HPC system
to run CI jobs for sites like GitHub or GitLab. However, most HPC sites disallow users from
running jobs on behalf of external systems.

Consider the open-source CI model, where an unknown user (or at least a user who is
unknown to the HPC center) submits a pull request (PR) to a project. The PR triggers jobs
that build and test the changed code in cloud environments.
Developers and maintainers often want to instead trigger jobs on a set of HPC platforms.
While the cloud allows users to provision isolated virtual machines and even isolated
virtual networks for CI jobs, most HPC systems lack this level of isolation. HPC sites implement
security at the facility boundary, allowing only trusted users in.
Once on the system, all users can access the shared filesystem and can
connect to compute nodes over the cluster network. A privilege escalation in this
environment could give a user access to other users' files, which may be export-controlled
or otherwise sensitive. HPC security teams therefore disallow setting up CI
to run arbitrary code, as it opens the site to such attacks.

Running code from {\it protected} branches, e.g., the
maintainer-approved main branch of a popular open-source project, may be allowed, but doing
this loses the
benefits of testing changes {\it before} they are integrated into the project. When the
PR is still open, contributors are motivated to fix issues that come out in testing
because they want their changes to be merged. If fixes are made after the fact, it can
be very difficult to keep fast-moving projects working for HPC.

\subsection{Administrative and political challenges}

A number of administrative and political reasons can hinder
progress on solutions to the above problems.
First, HPC sites do not typically prioritize build
or test cycles, because this is perceived to reduce cycles available
for production science runs, which is typically their raison d'\^etre. CI jobs tend
to be small and numerous, as opposed to the
more traditional larger, longer-running HPC jobs, and queuing policies that support this type of
work are not well understood, especially for heavily utilized systems with mostly larger jobs.
When asked how many cycles are needed for testing, users often reply with large numbers and
the need to test at scale. Facilities are reluctant to provide any one project
with a large testing allocation.

The tradeoff between using cycles for testing and saving cycles on production code that
may fail is not easy to quantify, but if public CI systems are any benchmark, a large
fraction of the benefit of CI can be realized through short-running builds and smoke
tests. Cloud CI services impose strict limits on job runtimes and resource
usage---typically just a few hours and one or two CPUs per job. All but the
largest codes can be built and tested for correctness within this footprint, at least at
a coarse granularity. Providing separate queues with similar policies on HPC systems
would require only a small fraction of overall system CPU hours, and while this approach
would not detect bugs that only appear at massive scale, it would still prevent many
production cycles from being wasted.

Because HPC sites are very focused on production jobs and production job performance,
very little interest has emerged in the HPC community for compute or network
virtualization. This is unfortunate, because these technologies would provide the type
of resource flexibility needed to run isolated, secure CI jobs. Infiniband, the most
popular HPC network, has very limited support for traffic isolation (8 or so isolated
channels---not enough for thousands of users), and most HPC systems still run
applications on bare metal instead of in VMs. Meanwhile, clouds have developed very
lightweight, secure VM solutions (e.g., Amazon Web Service's Nitro hypervisor) with
almost no virtualization overhead.

Finally, HPC center leadership has limited understanding of modern development workflows.
It is difficult to grasp the extent to which open-source software has spread
throughout the scientific software ecosystem, the rate at which modern software is
developed, and the interdependence of packages. The idea that key science applications
rely on externally developed software, that helping external software projects test on
HPC machines could be {\it beneficial} to internal projects, and that many {\it
  internal} projects are actually hosted and developed externally still needs
socializing in order to broaden understanding of the needs of modern software
developers.

\section{POTENTIAL SOLUTIONS}

Building and testing software on HPC systems has always been hard, but some solutions
to the challenges presented above have recently begun to emerge.

\subsection{Wisdom of the crowds}

Systems like Spack~\cite{gamblin+:sc15} and EasyBuild~\cite{hoste+:pyhpc12} have made
building on HPC systems easier by crowd-sourcing institutional build knowledge. These
systems include curated repositories of build scripts that aggregate and preserve
institutional knowledge of different machine environments and make HPC software easier
to build. While the projects themselves require extensive CI, changes are only checked
automatically with cloud CI, not on a diverse set of HPC resources. Without immediate,
automated testing of contributions, builds still frequently break and tests still frequently
fail in these environments.

\subsection{Jacamar and secure CI}

For {\it internal}, trusted projects at HPC centers, projects like Jacamar CI~\cite{jacamar-ci}
solve some of the security problems. They allow users to run CI jobs {\it as themselves} on HPC
machines, preserving the OS-level security boundaries that HPC centers require users to
adhere to. With Jacamar, one user cannot access and steal another user's data through the CI
system.
While internal projects {\it can} pull in trusted versions of external
software (e.g., recent releases), integration testing is still difficult. Internal
teams cannot easily test changes from PRs, because the changes in a PR cannot be
attributed to any trusted HPC center user. Without the ability to test PRs, incompatibilities
or bugs can be introduced through dependencies. Either the site must attribute every
PR to a known user, which is often not possible, or they must isolate the untrusted code
in its own environment.

\subsection{Separate resources for CI}

HPC sites are considering setting up separate resources for open-source CI. One of the
authors has been involved in such an effort at LLNL, to set up an isolated cluster {\it
  without} sensitive data, where public CI jobs can run with little risk to the main
HPC resources. The challenge with this approach is that it
duplicates effort---HPC system administrators are a scarce resource, and
maintaining an additional machine in a different network zone requires redundant work.
It is also difficult to ensure that the separate machine stays up to date with the main
systems.

\subsection{Vendor and cloud support}

As customers have come to rely on an increasing volume of open-source software, HPC vendors have
shown more interest in ensuring that this software works well on the platforms
they offer.  At the same time, more cloud vendors are producing their own HPC
offerings, and users can easily set up clusters in the cloud to run
HPC jobs. These clusters can even use a wide range of resource managers, like SLURM and PBS.
Such environments are not free, of course.

HPC vendors {\it may} begin to provide free, public cloud CI resources that open-source
developers could use to test their software. For large projects like
Spack, cycles can be donated in one place, but scaling the approach to support the many smaller,
independent HPC development projects that need CI is a much larger effort that
requires more cooperation between major HPC vendors and cloud platforms.

\subsection{Containerized environments}

In lieu of hardware resources, HPC vendors could also begin to provide containerized
versions of their software stack for building and testing in CI. Some vendors
have begun testing this approach, for example with Cray's Containerized Programming Environment
(CPE). Unfortunately, the container is currently only licensed to run on HPC resources,
so its versatility for build and test use cases is limited. It cannot be run in the
cloud, where adequate isolation and cycles are available.

For commodity clusters, it may be easier to provide containerized reproductions of the
production HPC environment, as there are not as many licensing issues involved.
Since most HPC sites are still administered very manually, HPC administrators will need
to lean into a culture of automation. If the site can provision the production environment
automatically, they can reliably provide a container with {\it exactly} the
same software that the main HPC site runs.

The idea of building containers based on production HPC environments is not new; one
of this paper's authors proposed it with his colleagues to NSF's XD
solicitation~\cite{NSF-XD} in 2008.  The proposal ultimately became part of NSF's
XSEDE environment, which operated between 2011 and 2022, but the containerization
component did not make it into the final, funded project.

More recently, NASA Ames~\cite{ames-hybrid-cloud} has successfully provisioned cloud-bursting
capabilities allowing users to build, test, and run codes in small allocations in cloud
environments before running them unmodified on the production Pleiades HPC cluster.
They leverage portable container workflows and abstract differences between cloud and
onsite resources through the MPI interface. This pioneering effort to create
``reproducible'' HPC infrastructure still has security limitations. Even in NASA's
environment, the cloud resources are provisioned in the same logical network as HPC
onsite resources. They cannot run untrusted code without risk to onsite data.

\subsection{More virtualization and IaaS}

The solution that likely makes most sense for HPC CI is to move towards a less trusting,
more isolated security model that would allow HPC systems to function more like clouds.
Flexible, isolated allocations for either internal or external CI jobs would
eliminate the duplication of effort required by many of the potential
approaches mentioned above. Isolated allocations also enable Infrastructure as a Service (IaaS)
within the HPC site, which would allow sites to mock entire distributed resource manager
environments and services. With this capability, developers could test entire workflow systems
in much more realistic scenarios.

We will need to work with vendors to develop and provide HPC environments with network
and OS support for isolation. In general, the only integrators currently providing these
capabilities widely are clouds, and there are {\it not} good on-premises solutions. HPC
sites will need to either start working with clouds more closely, or push
HPC vendors to provide like capabilities for HPC centers. It will likely be a long time
before truly ``converged'' infrastructure becomes widespread.

\section{CONCLUSION}

Continuous integration is indispensable for most software development and
maintenance today, including for scientific software.
However, CI is difficult to implement in HPC environments for many reasons.
Limitations of on-premises infrastructure preclude many of the security isolation
techniques used in modern cloud environments. HPC security policies must respect these
limitations by restricting the automation needed for responsive CI. Current solutions
require duplicated effort, either in provisioning dedicated resources for CI, or by
duplicating deployment effort with containerized environments. The most promising
solution is to move toward more automated, secure, flexible infrastructure,
which will be
neither quick nor easy to implement with the restrictions of today's HPC environment.

\section{ACKNOWLEDGMENT}

This article was inspired in part by a panel on Build, Integration and Testing for
Sustainable Scientific Computing Software, chaired by Keita Teranishi and Roscoe A.
Bartlett, at the 2022 SIAM Conference on Parallel Processing for Scientific Computing.
The authors thank the chairs for inviting us to the panel and creating the opportunity
for discussion. Daniel S. Katz thanks Ben Clifford and other members of the Parsl and
funcX teams for some of the ideas here. Part of this work was performed under the
auspices of the U.S. Department of Energy by Lawrence Livermore National Laboratory
under contract DE-AC52-07NA27344. Lawrence Livermore National Security, LLC
(LLNL-JRNL-846623).

\bibliographystyle{IEEEtran}
\bibliography{cise-hpc-ci}

\bigskip

\textbf{Todd Gamblin} received a BA degree in computer science and Japanese from Williams College, and the MS and PhD degrees in computer science from UNC Chapel Hill. He is a Distinguished Member of Technical Staff in the Livermore Computing division at Lawrence Livermore National Laboratory. His research interests include dependency management, software engineering, parallel computing, performance measurement, and performance analysis.
Contact him at tgamblin@llnl.gov.

\textbf{Daniel S. Katz} is Chief Scientist at the National Center for Supercomputing Applications, and a Research Associate Professor in computer science, electrical and computer engineering, and information sciences with the University of Illinois Urbana Champaign, Urbana, IL, USA. He received the Ph.D. degree in electrical engineering from Northwestern University, Evanston, IL, USA. His research interests include software applications, algorithms, fault tolerance, and programming in parallel and distributed computing, and policy issues, including citation and credit mechanisms and practices associated with software and data, organization and community practices for collaboration, and career paths for computing researchers.
He is a senior member of the IEEE, the IEEE Computer Society, and ACM, co-founder and current Associate Editor-in-Chief of the Journal of Open Source Software, co-founder of the US Research Software Engineer Association (US-RSE), and co-founder and steering committee chair of the Research Software Alliance (ReSA).
Contact him at d.katz@ieee.org.

\end{document}